\newenvironment{sketch}{\noindent{\it Proof Sketch:}\hspace*{1em}}{\qed
\medskip}
\newenvironment{Proof}{\noindent{\it Proof:}\hspace*{1em}}{\qed
\medskip}
\newcommand{\qed}{\rule{6pt}{6pt}}

\documentclass{llncs}
\usepackage{latexsym,graphics,epsfig,url}
\usepackage{amsfonts,amssymb}
\title{\bf Simultaneous Embedding of a Planar Graph \\and Its Dual on the Grid}
\author{Cesim Erten
\and Stephen G.~Kobourov
}
\institute{Department of Computer Science\\
University of Arizona\\
{\tt \{cesim,kobourov\}@cs.arizona.edu}
}

\begin{document}
\date{}
\maketitle

\thispagestyle{plain}
\begin{abstract}
Traditional representations of graphs and their duals suggest the
requirement that the dual vertices be placed inside their
corresponding primal faces, and the edges of the dual graph cross only
their corresponding primal edges. We consider the problem of
simultaneously embedding a planar graph and its dual into a small
integer grid such that the edges are drawn as straight-line segments
and the only crossings are between primal-dual pairs of edges. We
provide a linear-time algorithm that simultaneously embeds a 3-connected
planar graph and its dual on a $(2n-2)\times (2n-2)$ integer grid,
where $n$ is the total number of vertices in the graph and its
dual. Furthermore our embedding algorithm satisfies the two natural
requirements mentioned above. 

\end{abstract}

\section{ {\bf Introduction}}
In this paper we address the problem of simultaneously drawing a planar graph
 and its dual on a small
integer grid. The {\it{planar dual}} of an embedded planar graph $G$
is the graph $G'$ formed by placing a vertex inside each face of $G$,
and connecting those vertices of $G'$ whose corresponding faces in $G$
share an edge. Each vertex in $G'$ has a corresponding primal face and
each edge in $G'$ has a corresponding primal edge in the original
graph $G$. The traditional manual representations of a graph and its
dual, suggest two natural requirements. One requirement is that we
place a dual vertex inside its corresponding primal face and the
other is that we draw a dual edge so that it only crosses its
corresponding primal edge. We provide a linear-time algorithm that
simultaneously draws a planar graph and its dual using straight-line
segments on the integer grid while satisfying these two requirements.

Straight-line embedding a planar graph $G$ on the grid, i.e., mapping
the vertices of $G$ into a small integer grid such that each edge can
be drawn as a straight-line segment and that no crossings between
edges are created, is a well-studied graph drawing problem. The first
solution to this problem was given by Fraysseix, Pach and
Pollack~\cite{fpp-sssfe-88} who provided an algorithm that embeds a
planar graph on $n$ vertices on the $(2n-4)\times (n-2)$ integer
grid. Later, Schnyder~\cite{s-epgg-90} developed another method that
reduces the grid size to $(n-2)\times (n-2)$. Since then there have
been many studies regarding different restrictions of the
problem. Harel and Sardas~\cite{harel98algorithm} provide an algorithm
to embed a biconnected graph on a $(2n-4)\times (n-2)$ grid without
triangulating the graph initially. The algorithm of Chrobak and
Kant~\cite{chrobak97convex} embeds a 3-connected planar graph on a
$(n-2)\times (n-2)$ grid so that each face is convex. Miura, Nakano,
and Nishizeki~\cite{nakano4connected01} further restrict the graphs
under consideration to be 4-connected and present an algorithm for
straight-line embedding of such graphs on a $(\lceil n/2\rceil
-1)\times (\lfloor n/2\rfloor)$ grid.

Another related problem is that of simultaneously embedding more than
one planar graph. In particular, consider two planar graphs on the
same set of vertices: $H_1=(V,E_1)$ and $H_2=(V,E_2)$.  We would like
to embed $H_1$ and $H_2$ simultaneously so that the vertices in $V$
are mapped on the integer grid and each of $H_1$ and $H_2$ is realized
with straight-line segments and no crossings.  Similarly, we would like to
simultaneously embed two related graphs, not necessarily on the same
vertex set.  Such simultaneous embedding would enhance the visual
comparison of two graphs. In this paper we address the related problem
of embedding a planar graph and its dual on a small grid. Previous
researchers have considered two versions of the problem. 

In a paper
dating back to 1963, Tutte~\cite{t-hdg-63} shows that there exists a
simultaneous straight-line representation of a planar graph and its
dual in which the only intersections are between corresponding
primal-dual edges. However, a disadvantage of this representation is that
the area required by the algorithm can be exponential in the number of vertices of the graph. 
Bern and Gilbert~\cite{Bern:1992:DPD} address a
variation of the problem: finding suitable locations for dual
vertices, given a straight-line planar embedding of a planar graph, so
that the edges of the dual graph are also straight-line segments and cross
only their corresponding primal edges. They present a linear time
algorithm for the problem in the case of convex 4-sided faces and
show that the problem is NP-hard for the case of convex 5-sided
faces. 

In this paper we consider the problem of embedding a given
planar graph $G$ and its dual graph simultaneously so that following conditions are
met:
\begin{itemize}
\item The primal graph is drawn with straight-line segments without crossings.
\item The dual graph is drawn with straight-line segments without crossings.
\item Each dual vertex lies inside its primal face.
\item A pair of edges cross if and only if the edges are a primal-dual pair.
\item Both the primal and the dual vertices are on the $(2n-2)\times (2n-2)$ grid, 
where $n$ is the number of vertices in the primal and dual graphs.
\end{itemize}
In the next section we present a linear-time algorithm for this
problem which relies on finding a strictly convex drawing for fully
quadrilateralated graphs. 
\begin{figure}[t]
\begin{center}
\includegraphics[width=10cm]{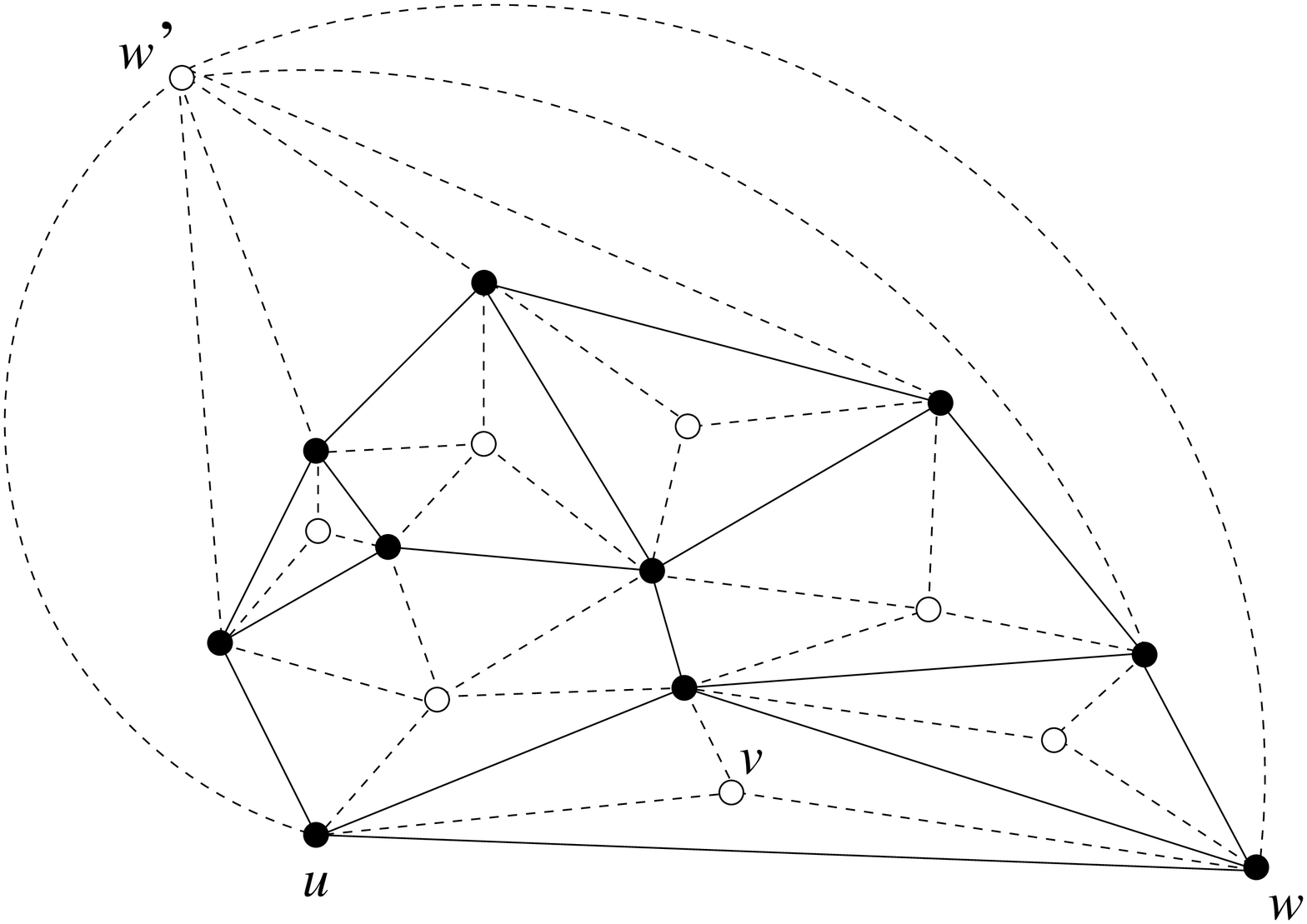}
\end{center}
\vspace{-.5cm}
\caption{\small 
3-connected graph $G_1$. The inserted dual vertices are shown as empty
circles. Dashed lines represent the inserted edges. To obtain $G_2$ we
remove the original edges of $G_1$ (drawn with solid lines).}
\label{construction}
\end{figure}

\vspace{-.3cm}
\section{ {\bf Algorithm for Embedding a Graph and Its Dual}}
Let $G_1$ be a 3-connected planar graph. We construct a new graph
$G_2$ that combines information about both the planar graph $G_1$ and
its dual. For this construction we make some changes in $G_1$. We
introduce a new vertex ${v_{i}}'$ corresponding to a face
${\mathcal{F}_{i}}'$ of $G_1$, for all $1\leq i \leq f$, where $f$ is
the number of faces of $G_1$. We connect each newly added vertex ${v_{i}}'$ to
each vertex $v_j$ of ${\mathcal{F}_{i}}'$ with a single new edge and
delete all the edges that originally belonged to $G_1$. Fig.~\ref{construction} 
shows a sample construction. We call the resulting planar graph $G_2$ 
{\it{fully-quadrilateralated(FQ)}},
i.e., every face of $G_2$ is a quadrilateral.
Since the original graph $G_1$ is 3-connected, the resulting graph $G_2$ is also
3-connected (proven formally in~\cite{t-hdg-63}). 


\vspace{.2cm}
{\bf Observation:} If we can embed the graph $G_2$ on the grid so that
each inner face of $G_2$ is strictly convex and the outer face of
$G_2$ lies on a strictly concave quadrilateral, then we can embed the
initial graph $G_1$ and its dual so that we meet all the problem
requirements with the only exception that one edge of the primal graph
$G_1$(or its dual) is drawn with one bend.

\vspace{.2cm}
The requirement that the edges of the dual graph be straight and cross
only their corresponding primal edges is guaranteed by the strict
convexity of the quadrilateral faces. Let the outer face of the graph
$G_2$ be $(u, v, w, w')$, where $u, w$ are primal vertices and $v, w'$
are dual vertices, as shown in Fig.~\ref{construction}. The exception
arises from the fact that we need to draw $(u,w)$ and
$(v,w')$, while both of these edges can not lie inside the quadrilateral
$(u, v, w, w')$. In order to get around this problem we embed the
quadrilateral $(u, v, w, w')$ so that it is strictly concave. This way
only one bend for one of the edges $(u,w)$ or $(v,w')$ will be
sufficient. As a result all the edges in the primal and the dual graph
are embedded as straight-lines, except for one edge. In fact, it is
easy to choose the exact edge we need (either from the primal or from
the dual).

Hence, the original problem can
be transformed into a problem of straight-line embedding an FQ-3-connected 
planar graph $G$ on the grid so that
each internal face of $G$ is strictly convex and the outer face of $G$ lies on a strictly 
concave quadrilateral. Note that this problem can be solved
by the algorithm of Chrobak {\em et al.}~\cite{cgt-cdgtt-96}. However, the area guaranteed by their algorithm
is $O(n^3)\times O(n^3)$, whereas our algorithm
guarantees a
drawing on the $(2n-2)\times (2n-2)$ grid, which is stated in the main theorem in this paper:

\begin{theorem} 
Given a 3-connected planar graph $G_1$, we can embed $G_1$ and its dual on
a $(2n-2)\times(2n-2)$ grid, where $n$ is the number of vertices in $G_1$ and 
its dual, so that each dual vertex lies inside its primal face, 
each dual edge crosses only its primal edge and every edge in the overall
embedding is a straight-line segment except for one edge which has a bend placed 
on the grid. Furthermore, the running time of the algorithm is $O(n)$.
\end{theorem} 

\vspace{-.3cm}
\subsection{ {\bf Overview of the Algorithm}}
Given a 3-connected graph $G_1$, we summarize our algorithm to simultaneously 
embed $G_1$ and its dual as follows:

\vspace{.2cm}
$\bullet$ Find a topological embedding of $G_1$ using~\cite{ht-ept-74}. 

\vspace{.1cm}
$\bullet$ Apply the construction described above to find $G_2$. 

\vspace{.1cm}
$\bullet$ Let $G=G_2$, where $G$ is an FQ-3-connected planar graph. 

\vspace{.1cm}
$\bullet$ Find a suitable canonical labeling of the vertices of $G$.

\vspace{.1cm}
$\bullet$ Place the vertices of $G$ on the grid one at a time using this ordering. 

\vspace{.1cm}
$\bullet$ Remove all the edges of $G$ and draw the edges of $G_1$ and its dual. 

\vspace{.2cm}
Note that our method works only for 3-connected graphs. A commonly
used technique for drawing a general planar graph is to embed the
graph after fully triangulating it by adding some extra edges and then
to remove the extra edges from the final embedding. Using the same
idea, we could first fully triangulate any given planar graph. Then
after embedding the resulting 3-connected planar graph and its dual,
we could remove the extra edges that were inserted initially.
However, the problem with this approach is that after removing the
extra edges there could be faces with multiple dual vertices
inside. Thus the issue of choosing a suitable location for the duals
of such faces remains unresolved. In fact, depending on the drawing of
that face, it could as well be the case that no suitable location for
the dual exists~\cite{Bern:1992:DPD}. In the rest of the paper we
consider only 3-connected graphs.

\vspace{-.3cm}
\subsection{ {\bf The Canonical Labeling}}
We present the canonical labeling for the type of graphs under
consideration.  It is a simple restriction of the canonical labeling
of~\cite{k-dpguc-96}, which in turn is based on the ordering
defined in~\cite{fpp-sssfe-88}.

Let $G$ be an FQ-3-connected planar graph with $n$ vertices. Let $(u, v, w, w')$
be the outer face of $G$ s.t. $u, w$ are primal vertices and $v, w'$ are dual vertices. 
Then there exists a mapping $\delta$ from the vertices of $G$
onto $v_i$, $1\leq i\leq m$ such that $\delta$ maps $u$ and $v$ to
$v_1$, $w'$ to $v_m$ and satisfies the following invariants for every $3\leq k \leq
m$:
\hspace*{.5cm}
\begin{enumerate}
\item 
The subgraph $G_{k-1}\subseteq G$, induced by the vertices labeled
$v_i$, $1\leq i\leq k-1$ is biconnected and the boundary of its
exterior face is a cycle $C_{k-1}$ containing the edge $(u, v)$.
\item 
Either one vertex or two vertices can be labeled $v_k$.
\begin{enumerate}
\item 
Let $z_0$ be the only vertex labeled $v_k$. Then $z_0$ belongs to the
exterior face of $G_{k-1}$, has at least two neighbors in $G_{k-1}$
and at least one neighbor in $G-G_k$.
\item 
Let $z_0, z_1$ be the two vertices labeled $v_k$, where $(z_0, z_1)$
is an edge in $G$. Then $z_0, z_1$ belong to the outer face of
$G_{k-1}$, each has exactly one neighbor in $G_{k-1}$ and at least
one neighbor in $G-G_k$.
\end{enumerate}
\end{enumerate}

\begin{figure}[t]
\begin{center}
\includegraphics[width=10cm]{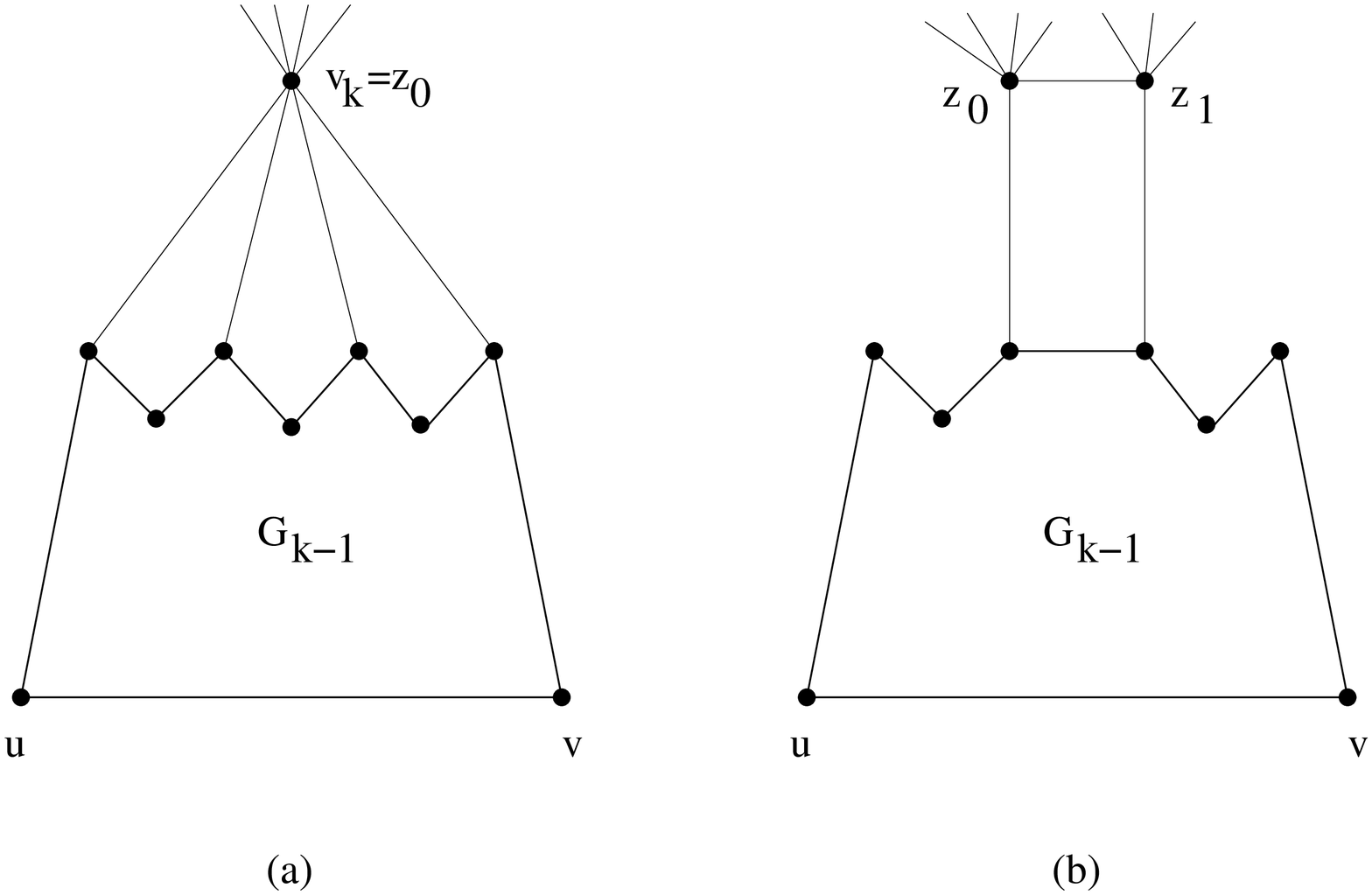}
\end{center}
\vspace{-.5cm}
\caption{\small a)Only one vertex, $z_0$, is labeled $v_k$    b)Two vertices, $z_0$ and $z_1$ are labeled $v_k$. 
}
\label{ordering}
\end{figure}

Since $G$ is $FQ$, all the faces created by adding 
$v_k$, $3\leq k\leq m$,  have to be quadrilaterals, see Fig.~\ref{ordering}.

Note that assigning the mappings onto $v_1$ and $v_m$ as above provides us the embedding where
all the edges of both the primal and the dual graph are straight except for one 
primal edge, $(u, w)$, which has a bend. Alternatively assigning $v$ and $w$ to map onto
$v_1$, and $u$ to map onto $v_m$ would choose a dual edge, $(v, w')$, to have a bend.
 
\begin{lemma} 
Every FQ-3-connected planar graph has a canonical
labeling as defined above.
\end{lemma}

Kant~\cite{k-dpguc-96} provides a linear-time algorithm to find a canonical
labeling of a general 3-connected planar graph. It is easy to see that 
the canonical labeling definition of~\cite{k-dpguc-96} when applied
to FQ-3-connected planar graphs, gives us the labeling defined above.  

\vspace{-.3cm}

\subsection{ {\bf The Placement of the Vertices}}
The main idea behind most of the straight-line grid embedding
algorithms is to come up with a suitable ordering of the vertices and
then place the vertices one at a time using the given order, while
making sure that the newly placed vertex (or vertices) is (are)
visible to all the neighbors. In order to realize this last goal, at
each step, a set of vertices are shifted to the right without
affecting the planarity of the drawing so far. Our placement algorithm
is similar to the algorithm of Chrobak and
Kant~\cite{chrobak97convex}, with some changes in the invariants that
we maintain to guarantee the visibility together with strict convexity
of the faces.

Let the canonical labeling, $\delta$, that maps the vertices of $G$
onto $v_1, v_2, ...v_m$ be defined as in the previous section. Let
$\mathcal{U}(g_i)$ denote the vertices {\it{under}}
$g_i$. $\mathcal{U}(g_i)$ should be shifted to the right whenever the
vertex $g_i$ is shifted to the right.  $\mathcal{U}(g_i)$ is
initialized to $\{g_i\}$ for every vertex $g_i$ of $G$.  Let
$\delta(g_i)=v_{i'}$ and $\delta(g_j)=v_{j'}$. Then we define
$Low(g_i, g_j)=i$ if $i'<j'$, $Low(g_i, g_j)=j$ if $j'<i'$. If $i'=j'$
then let $Low(g_i, g_j)$ be the one that is placed to the left. Let $x(g_i)$,
$y(g_i)$ respectively denote the $x$ and $y$ coordinates of the vertex $g_i$.

\vspace{.2cm}
{\it {$\bullet$Embed the First Quadrilateral Face:}}\hspace*{.2cm}
We start by placing the vertices mapped onto $v_1$ and $v_2$. The ones
that are mapped onto $v_1$ are $u$ and $v$. We place $u$ at $(0,0)$
and $v$ at $(3,0)$. Note that two vertices should be mapped to $v_2$. 
We place the vertex that is mapped to $v_2$ and that
has an edge with $u$ at $(1,1)$ and the other at $(2,1)$. 

\vspace{.2cm}
Then, for every k, $3\leq k\leq m$, we do the following:

\vspace{.2cm}
{\it {$\bullet$Update $\mathcal{U}(g_i)$:}}\hspace*{.2cm} Let
$C_{k-1}=(u=c_1, c_2, ..., c_r=v)$.  Let $c_p, c_q\in C_{k-1}$,
respectively be the first and the last neighbor of the
vertex(vertices) mapped to $v_k$.  If only one vertex, $z_0$, is
mapped to $v_k$, we update $\mathcal{U}(c_p)$, $\mathcal{U}(c_q)$ and
$\mathcal{U}(z_0)$ as follows:

\[Low(c_p, c_{p+1})=p+1 \Longrightarrow \mathcal{U}(c_p)=\mathcal{U}(c_p) \cup \mathcal{U}(c_{p+1})\]
\[Low(c_{q-2}, c_{q-1})=q-2 \Longrightarrow \mathcal{U}(c_q)=\mathcal{U}(c_{q}) \cup \mathcal{U}(c_{q-1})\]
\[\mathcal{U}(z_0)=\mathcal{U}(z_0)\cup \bigcup_{i=Low(c_{p},c_{p+1})+1}^{Low(c_{q-2},c_{q-1})}\mathcal{U}(c_i)\]

We do not change $\mathcal{U}(g_i)$ if two vertices, $z_0$ and $z_1$,
are mapped to $v_k$.

\vspace{.2cm}
{\it {$\bullet$Shift to the right:}}\hspace*{.2cm} We then perform the
necessary shifting.  We shift each vertex $g_i \in
\bigcup_{i=q}^{r}\mathcal{U}(c_i)$ to the right by one if only one
vertex is mapped to $v_k$, by two otherwise.

\vspace{.2cm}
{\it {$\bullet$Locate the New Vertices:}}\hspace*{.2cm} Finally we
locate the vertex(vertices) mapped to $v_k$ on the grid. Let
$|v_k|$ denote the number of vertices mapped to $v_k$. Then we have:

\vspace{.2cm}
If $c_p$ has no neighbors in $G-G_k$ \\
\hspace*{1cm}$x(z_0)=x(c_p)$ \\
\hspace*{1cm}$y(z_0)=y(c_q)+x(c_q)-x(c_p)-|v_k|+1$ \\
\hspace*{.5cm}otherwise \\
\hspace*{1cm}$x(z_0)=x(c_p)+1$ \\
\hspace*{1cm}$y(z_0)=y(c_q)+x(c_q)-x(c_p)-|v_k|$ \\
\hspace*{.5cm}If $|v_k|=2$ define $z_1$ also: \\
\hspace*{1cm}$x(z_1)=x(z_0)+1$ \\
\hspace*{1cm}$y(z_1)=y(z_0)$

\begin{figure}[t]
\begin{center}
\includegraphics[width=12cm]{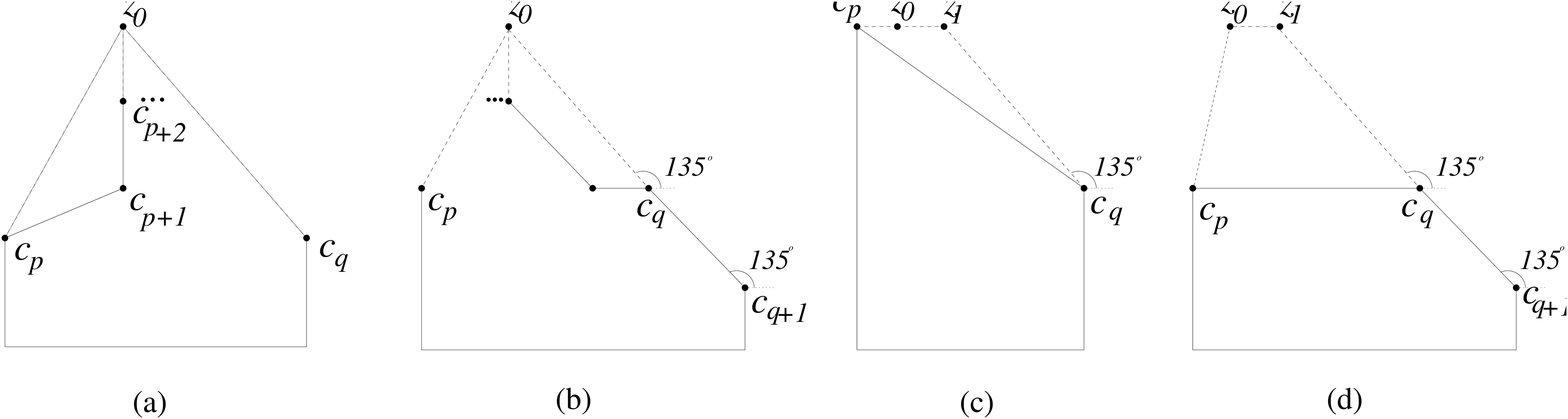}
\end{center}
\vspace{-.5cm}
\caption{\small Possible degenerate cases. a)Type $d_1$  b)Type $d_2$  
c)Type $d_3$  d)Type $d_4$  
}
\label{degeneracy}
\end{figure}

\vspace{.2cm}
Upto this step the algorithm is just a restriction of the one in~\cite{chrobak97convex}
and it guarantees the convex drawing of the faces. Then, in order to guarantee strict-convexity,
we note the following degenerate cases, see Fig.~\ref{degeneracy}: 

\vspace{.2cm}
{\it {$\bullet$Degeneracies:}}\hspace*{.2cm} We check for the following: 

\vspace{.2cm}
If only one vertex, $z_0$, is mapped to $v_k$ \\
\hspace*{.7cm} $^{\tiny{(d_1)}}$ If $x(z_0)=x(c_{p+1})=x(c_{p+2})$ \\
\hspace*{1.8cm} Shift each vertex $g_i \in \bigcup_{i=p+1}^{r}\mathcal{U}(c_i)$ to the right by one. \\
\hspace*{1.8cm} Perform the location calculation for $z_0$ again. \\
\hspace*{.7cm} $^{\tiny{(d_2)}}$ If $k<m$ and $z_0, c_q, c_{q+1}$ are aligned and
$c_q$ has no neighbors in $G-G_k$ \\
\hspace*{1.8cm} Shift each vertex $g_i \in \bigcup_{i=q+1}^{r}\mathcal{U}(c_i)$ to the right by one. \\
\hspace*{.5cm} If two vertices, $z_0$ and $z_1$ are mapped to $v_k$ \\
\hspace*{.7cm} $^{\tiny{(d_3)}}$ If $y(z_0)=y(z_1)=y(c_p)$ \\
\hspace*{1.8cm} Shift each vertex $g_i \in \bigcup_{i=q}^{r}\mathcal{U}(c_i)$ to the right by one. \\
\hspace*{1.8cm} Perform the
location calculation for $z_0$ and $z_1$ again. \\
\hspace*{.7cm} $^{\tiny{(d_4)}}$ If $k<m$ and $z_1, c_q, c_{q+1}$ are aligned and
$c_q$ has no neighbors in $G-G_k$ \\
\hspace*{1.8cm} Shift each vertex $g_i \in \bigcup_{i=q+1}^{r}\mathcal{U}(c_i)$ to the right by one. \\

\vspace{-.3cm}

\subsection{ {\bf Proof of Correctness}}
\begin{lemma} 
Let $C_{k}=(u=c_1, c_2, ..., c_r=v)$ be the exterior face of $G_k$
after the $k^{th}$ placement step. Let $\alpha(c_j, c_{j+1})$ denote
the angle of the vector $\vec{c_jc_{j+1}}$, for $1\leq j\leq r-1$. The
following holds for $2\leq k\leq m-1$:
\begin{enumerate}

\item 
$\alpha(c_j, c_{j+1})$ lies in $[-45^{\circ}, \arctan -1/2]
\cup \{0\} \cup [45^{\circ}, 90^{\circ}]$. It can not lie in $(-45^{\circ}, \arctan -1/2]$ if $c_j$ has a neighbor in $G-G_k$.

\item 
If $c_j\in C_{k}, c_j\notin \{c_1, c_r\}$ s.t. $c_j$ does not have a
neighbor in $G-G_k$, then:
\begin{enumerate}
\item If $Low(c_{j-1}, c_j)=j-1$ then $\alpha(c_j, c_{j+1})=90^{\circ}$ otherwise 
$\alpha(c_{j-1}, c_j)=-45^{\circ}$.
\item If $\alpha(c_j, c_{j+1})=90^{\circ}$ then $\alpha(c_{j-1}, c_j)\neq 90^{\circ}$.
\item If $\alpha(c_j, c_{j+1})=-45^{\circ}$ then $\alpha(c_{j-1}, c_j)\neq -45^{\circ}$. 
\end{enumerate} 
\end{enumerate}
\end{lemma}

We provide the proof of the above lemma in
the Appendix.  

\vspace{.2cm}
{\it { Preserving Planarity}}\hspace*{.2cm} Let only one vertex,
$z_0$, be mapped to $v_k$. If $(z_0, c_j)$ is an edge in $G_k$ for some $c_j \in C_{k-1}$,
then the placement algorithm and the previous lemma guarantees
that $-90<\alpha (z_0,c_j)<-45$, for $j\neq p, j\neq q$. 
Then no crossing is
created between a new edge $(z_0, c_j)$ and the edges of
$C_{k-1}$. Because such a crossing would imply that there exists
$j'<j$ s.t.  $c_{j'}\in C_k$ and $\alpha(c_{j'}, c_j)<-45$. But this
is impossible by the first part of the above lemma.  The same idea
applies to the case where $|v_k|=2$. Then the following corollary
holds:

\begin{corollary}Insertion 
of the vertex(vertices) mapped to $v_k$, at the $k^{th}$ placement
step, where $2\leq k\leq m$ preserves planarity.
\end{corollary}

\begin{figure}[t]
\begin{center}
\includegraphics[width=12cm]{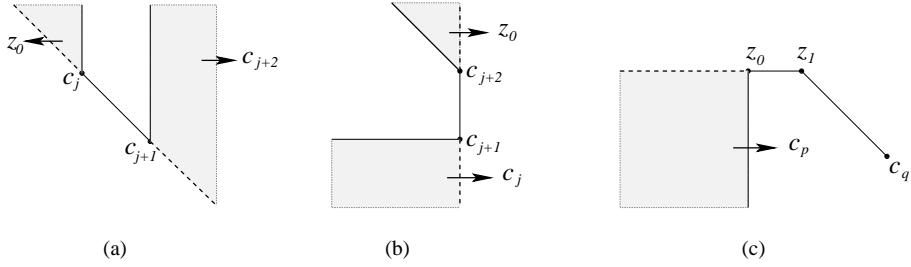}
\end{center}
\vspace{-.5cm}
\caption{\small 
The vertices pointed to by the arrows must lie in the indicated
area. The dashed lines are to indicate open boundaries that are not
included in the area.  }
\label{convexity}
\end{figure}

\vspace{.2cm}
{\it{Strictly Convex Faces}}\hspace*{.2cm} Let $|v_k|=1$ and $z_0$ be
the vertex mapped to $v_k$.  Let $\mathcal{F}_j=(c_j, c_{j+1},
c_{j+2}, z_0)$ be a quadrilateral face created after the insertion of
$z_0$.  If $Low(c_{j}, c_{j+1})=j+1$, then by the previous lemma
$\alpha(c_j, c_{j+1})=-45^{\circ}$.  Fig.~\ref{convexity}(a) shows the
area where $z_0$ and $c_{j+2}$ must lie.  If $Low(c_{j}, c_{j+1})=j$,
then $\alpha(c_{j+1}, c_{j+2})=90^{\circ}$.  Fig.~\ref{convexity}(b)
shows the area where $z_0$ and $c_{j+2}$ must lie in this case. Both
cases imply that $\mathcal{F}_j=(c_j, c_{j+1}, c_{j+2}, z_0)$ is
strictly convex.

If $|v_k|=2$ and $z_0, z_1$ are mapped to $v_k$, the placement
algorithm requires that $c_p$ must lie in the area shown in
Fig.~\ref{convexity}(c), which implies that the newly created face is
strictly convex.  The following corollary holds:

\begin{corollary}The newly created faces after the insertion 
of the vertex(vertices) mapped to $v_k$, at the $k^{th}$ placement
step, where $2\leq k\leq m$, are strictly convex.
\end{corollary} 

\vspace{.2cm}
{\it {Shifting Preserves Planarity and Strictly Convex
Faces}}\hspace*{.2cm} The above discussion shows that after the
insertion of the vertex(vertices) at the $k{th}$ placement step, no
new edge crossing is created and all the newly added faces are
strictly convex. In order to complete the proof of correctness we only
need to prove that the same holds for shifting also:

\begin{lemma} 
Let $C_{k}=(u=c_1, c_2, ..., c_r=v)$ be the exterior face of $G_k$
after the $k^{th}$ placement step, where $2\leq k<m$. For any given $j$, 
where $1\leq
j\leq r$, shifting the vertices in
$\bigcup_{i=j}^{r}\mathcal{U}(c_i)$, to the right by $s$ units
preserves the planarity and the strictly convex faces of $G_k$.
\end{lemma}
\begin{sketch}
The claim holds trivially for $k=2$. Assume it holds for $k'=k-1$,
where $2\leq k'< m-1$. We assume $|v_k|=1$. The case where $|v_k|=2$
is similar. Let $z_0$ be the vertex mapped to $v_k$ and $c_p, c_q \in
C_{k-1}$, respectively be the first and the last neighbor of $z_0$in
$G_{k-1}$.

If $j\leq p$ then by the inductive assumption the planarity of
$G_{k-1}$ and the strictly convex faces of $G_{k-1}$ are
preserved. The faces introduced by $z_0$ shifts rigidly to the right,
which, by the previous corollaries, implies that $G_k$ is planar and
all its faces are strictly convex.

If $j>q$, then by the inductive assumption the planarity of $G_{k-1}$
and the strictly convex faces are preserved. Since neither $z_0$ nor
any of its neighbors in $G_{k-1}$ are shifted the lemma follows.

If shifting the newly inserted vertex $z_0$, we inductively apply the
shifting to $j'=Low(c_p, c_{p+1})+1$ in $G_{k-1}$.  By the inductive
assumption the planarity and strictly convex faces are preserved for
$G_{k-1}$. Since we applied a shifting starting with $j'$ then, all
the faces except the first one are shifted rigidly to the right, which
implies that those faces are strictly convex. Then the only problem
could arise with the leftmost face. If $Low(c_p, c_{p+1})=p$, then
$c_{p+1}, c_{p+2}$ and $z_0$ are all shifted to the right by the same
amount. Since initially the face $(c_p, c_{p+1}, c_{p+2}, z_0)$ was
strictly convex, it continues to be so after shifting those three
vertices also. In the case where $Low(c_p, c_{p+1})=p+1$, the only
shifted vertices are $z_0$ and $c_{p+2}$. Again shifting those two
vertices does not change the property that the face is convex.

If $j=q$, the situation is very similar to the previous case, except
now the only deformed face is the rightmost face, instead of the
leftmost one. The same idea applies to this case also, i.e., given
that initially the face is strictly convex, it remains to be so after
shifting
\end{sketch}

\vspace{-.3cm}
\subsection{ {\bf Grid Size}}
\begin{lemma} The placement algorithm requires a grid of size at most
$(2n-4)\times(2n-4)$.
\end{lemma}
\begin{sketch} If no degeneracies are created then the exact grid size required
is $(n-1)\times(n-1)$. We show that each degenerate case can be associated with a
newly added quadrilateral face of $G$. 

Degenerate case of type $d_1$ is associated with the 
face $(c_p,c_{p+1},c_{p+2},z_0)$. Degenerate case of type $d_2$ at some step $k$ of the algorithm, 
is associated with a face $(z_0,c_q,c_{q+1},g_i)$, where $g_i$ is a vertex that will be added at some
step $k'>k$ of the algorithm. We know that such a face exists, since $k<m$, $c_q$ has no neighbors
in $G-G_k$ and each face under consideration is a quadrilateral. Similar argument holds for degenerate 
case of type $d_4$. Finally degenerate case of type
$d_3$ is associated with the face $(c_p,c_q,z_1,z_0)$. Fig.~\ref{degeneracy} shows all four 
types of degeneracies that can occur. Note that each quadrilateral face is 
associated with at most one degeneracy. 

Since an {\it {FQ}} graph
$G$ with $n$ vertices has $n-3$ inside faces, 
the placement algorithm requires grid size of at most $(2n-4)\times(2n-4)$.
\end{sketch}  

\vspace{.2cm}
{\it { Final Shifting}}\hspace*{.2cm} Let $(u, v, w, w')$ be the outer face of $G$. The 
placement algorithm and Lemma-2 imply that the outer face is the 
isosceles right triangle $\triangle{uvw'}$ and 
that $w$ lies on the line segment $(v,w')$. We need to do one final right shift 
to guarantee that the outer face $(u, v, w, w')$ lies on a strictly concave quadrilateral.
For this we just shift $v$ to the right by one. As a result we can draw 
the edge $(v, w')$ as a straight-line segment. In order to draw the edge
$(u, w)$, we place a bend point at $(x(w')-1, y(w')+2)$, where $x(w')$ and $y(w')$,
respectively denote the $x$ and $y$ coordinates of the vertex $w'$. We connect the bend 
point with $u$ and $w$. Then the total area required is $(2n-2)\times (2n-2)$ and 
Theorem-1 follows. 

\vspace{-.3cm}
\section{Implementation}
\begin{figure}[t]
\begin{minipage}[b]{12cm}
\begin{center}
\includegraphics[width=3cm]{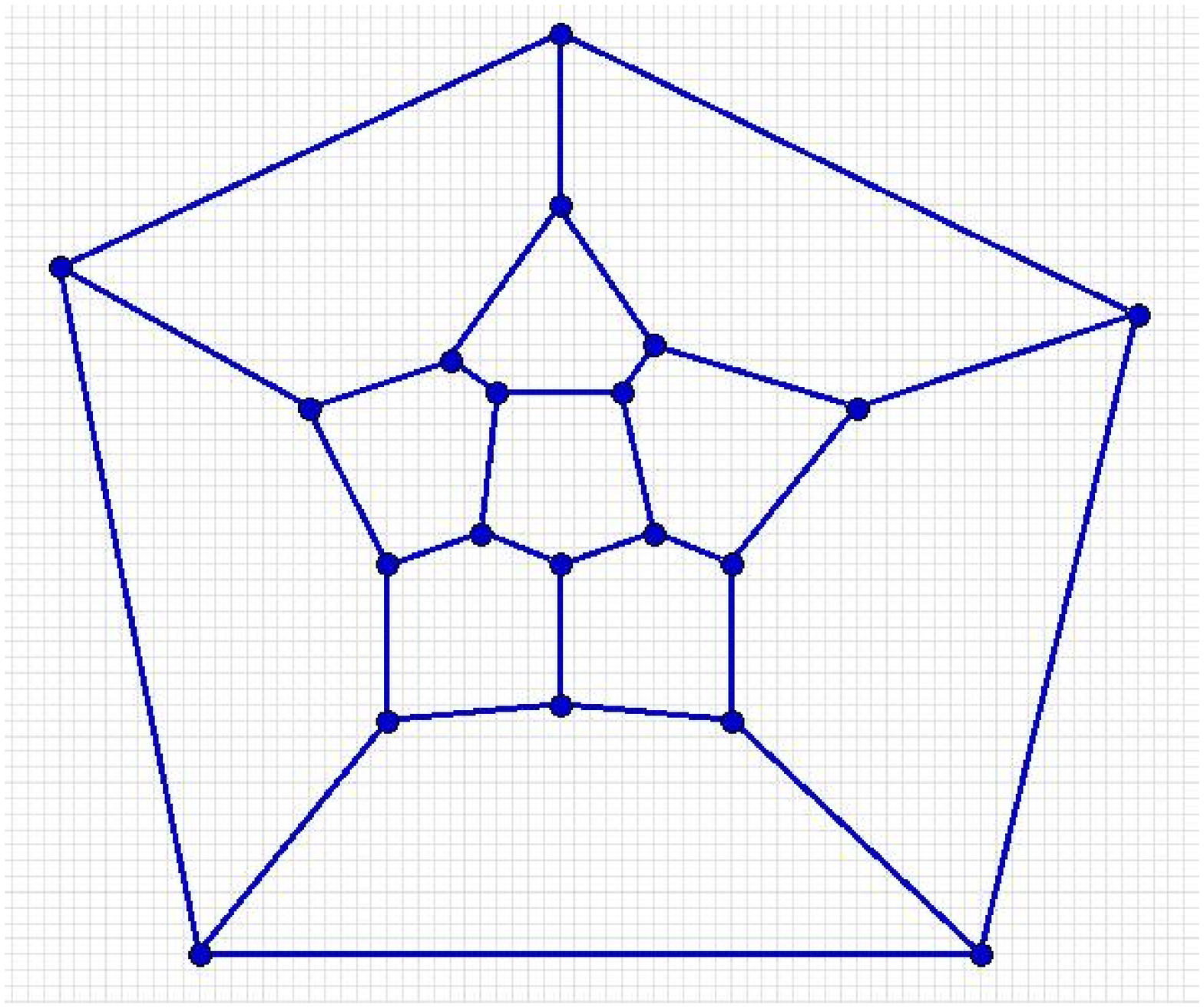}
\hspace{.1cm}
\includegraphics[width=7.5cm]{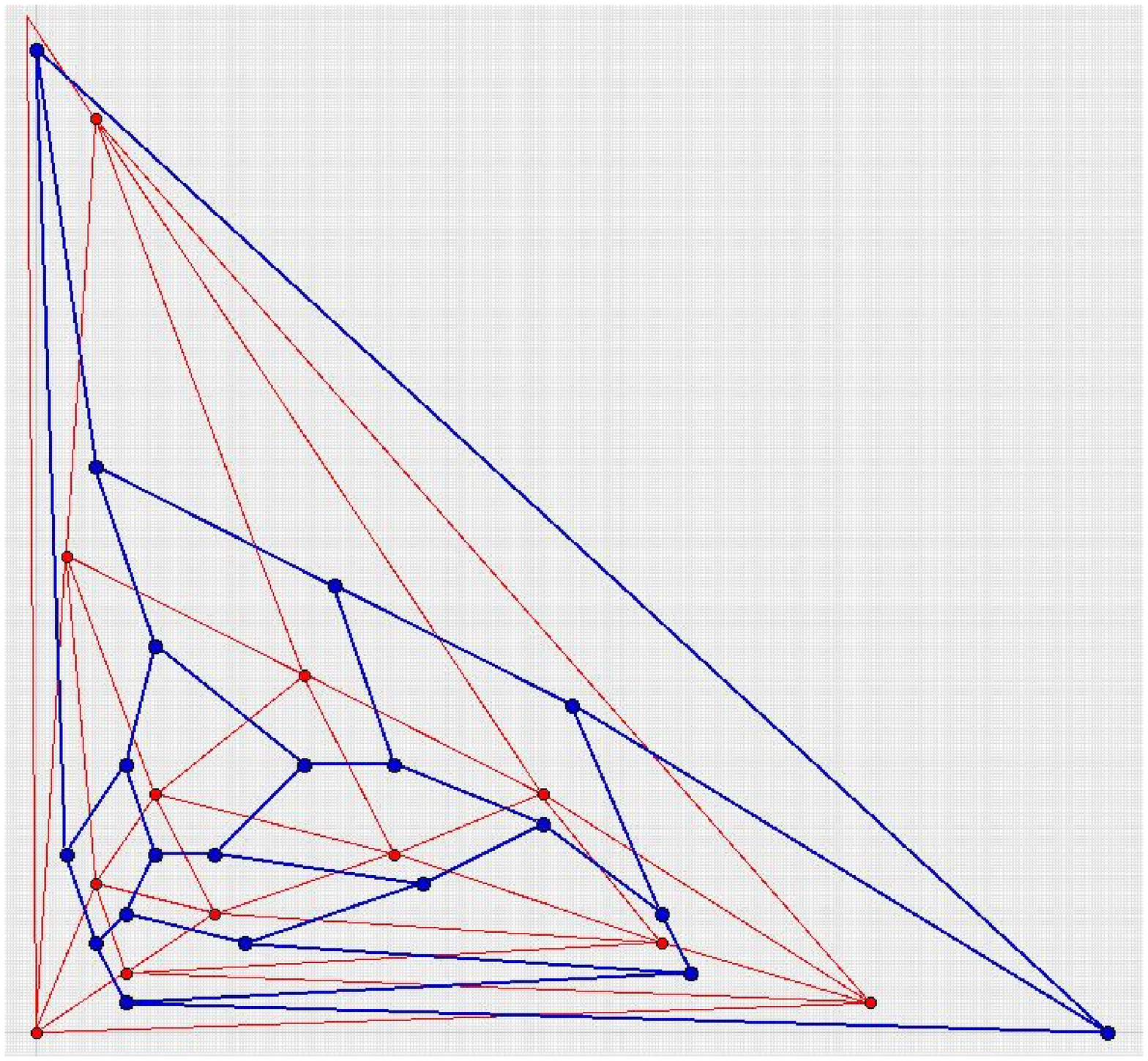}
\end{center}
\end{minipage}
\caption{\small\sf Dodecahedral graph and its dual representation.
The blue vertices(edges) in the primal/dual representation 
correspond to vertices(edges) of the primal graph, and the red ones correspond 
to the ones of the dual.}
\label{drawings1}
\end{figure}

\begin{figure}
\begin{minipage}[b]{12cm}
\begin{center}
\includegraphics[width=3cm]{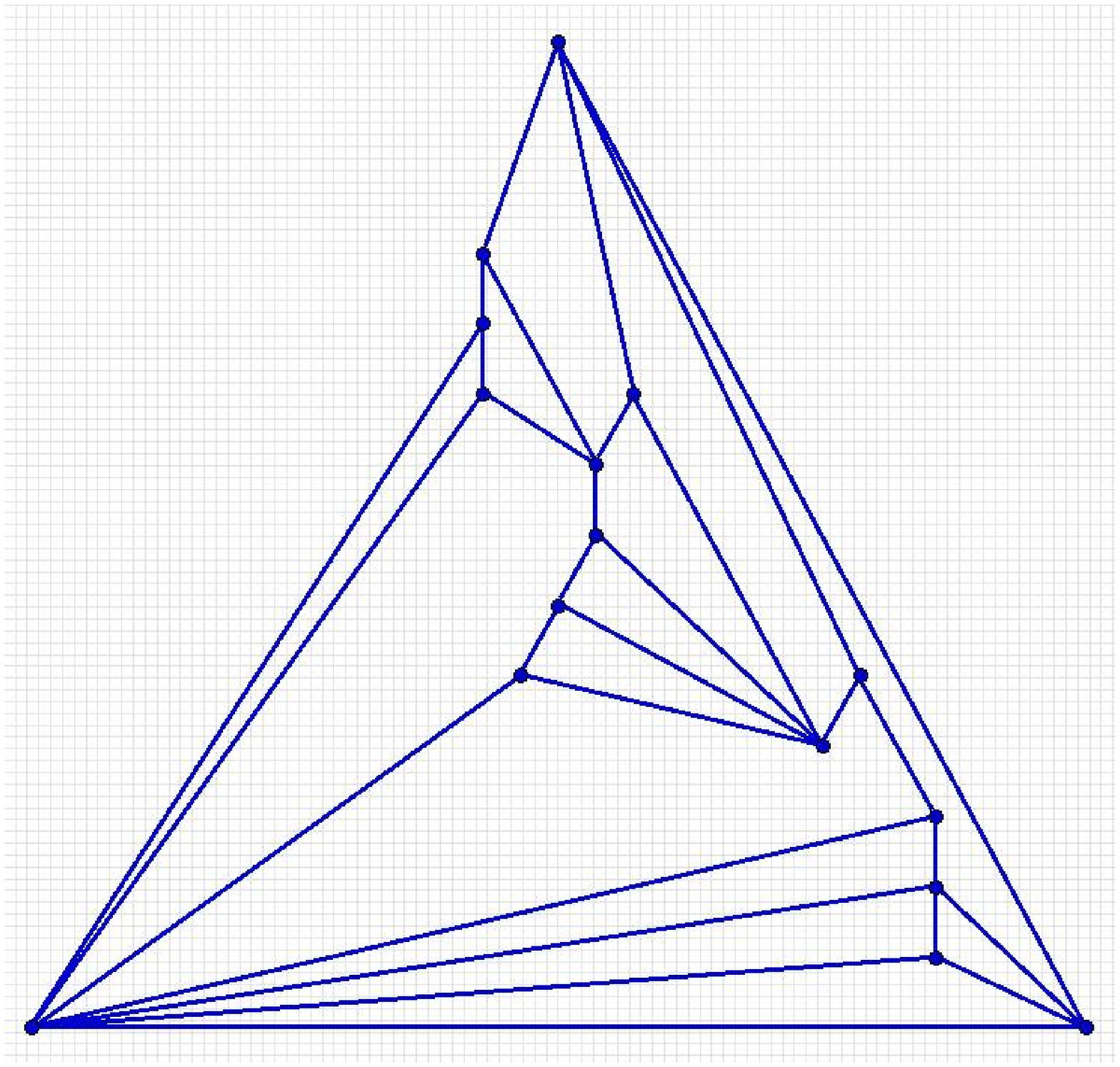}
\hspace{.1cm}
\includegraphics[width=7.5cm]{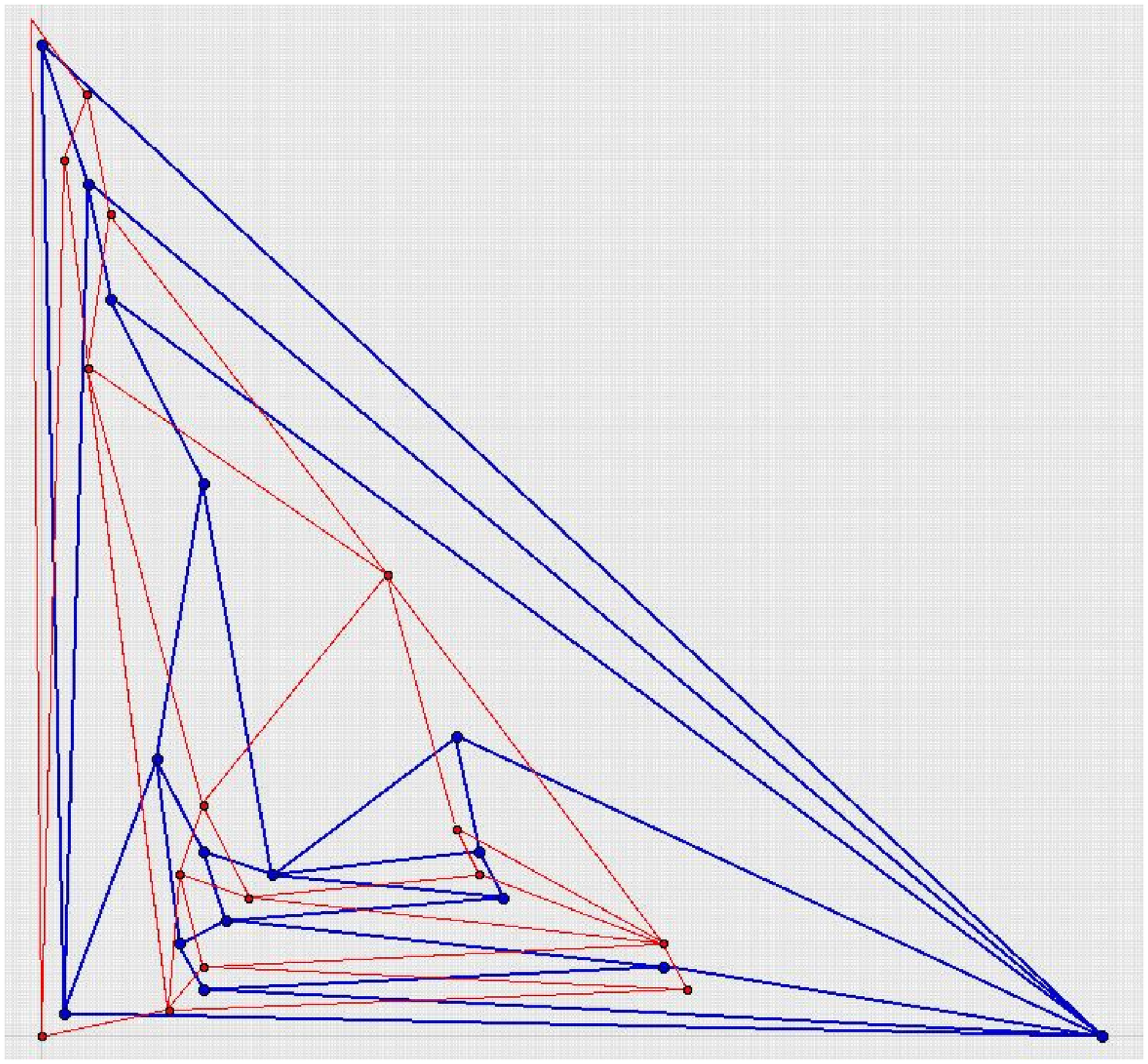}
\end{center}
\end{minipage}
\caption{\small\sf An arbitrary 3-connected planar graph with 16 vertices and its
dual representation.}
\label{drawings2}
\end{figure}

We have implemented our algorithm to visualize 3-connected planar graphs and their duals. 
Finding a suitable canonical labeling takes linear time~\cite{k-dpguc-96}. We make use 
of the technique introduced by~\cite{cp-ltadp-95} to do the placement step. It is based 
on the fact that storing relative x-coordinates of the previously embedded vertices is 
sufficient at every step. Then the placement step also requires only linear time. 
Overall, the algorithm runs in linear time.    
Fig.~\ref{drawings1} shows the primal/dual drawing we get for the dodecahedral graph and 
Fig.~\ref{drawings2} shows the primal/dual drawing of an arbitrary 
3-connected planar graph. 

\vspace{-.3cm}
\section{Conclusion and Open Problems}
We have shown how to embed a planar graph and its dual on a small grid
so that the embedding satisfies certain criteria. In particular, the dual 
vertices should be placed inside
their primal faces and the dual edges should cross only their primal
edges. We have provided a linear-time algorithm that finds a
straight-line planar embedding of a 3-connected planar graph and its
dual on a $(2n-2)\times (2n-2)$ grid such that the embedding satisfies
the requirements.

The following open problems arise from this work. Is there a larger
class of planar graphs that allows for
primal-dual embedding on a small grid, so that the drawing requirements
can be met?  For what class of planar graphs can we guarantee 
stronger results, such as perpendicular planar-dual crossing, i.e.,
one in which the dual edges cross the primal edges at right angles.
Finally, how can we generalize the idea of simultaneous embedding of
graphs not only for planar-dual pairs, but to any given two planar
graphs, so that the resulting embedding of the graphs provides a nice
representation and enhances the visual comparison between the two?

\vspace{-.3cm}
\section{Acknowledgements}
We would like to thank Anna Lubiw for introducing us to the problem of 
simultaneous graph embedding and for stimulating discussions about it.

\clearpage
{\small
\bibliographystyle{abbrv}
\bibliography{stephen.bib}

\begin{thebibliography}{10}

\bibitem{Bern:1992:DPD}
M.~Bern and J.~R. Gilbert.
\newblock Drawing the planar dual.
\newblock {\em Information Processing Letters}, 43(1):7--13, Aug. 1992.

\bibitem{cgt-cdgtt-96}
M.~Chrobak, M.~T. Goodrich, and R.~Tamassia.
\newblock Convex drawings of graphs in two and three dimensions.
\newblock In {\em Proc. 12th Annu. ACM Sympos. Comput. Geom.}, pages 319--328,
  1996.

\bibitem{chrobak97convex}
M.~Chrobak and G.~Kant.
\newblock Convex grid drawings of 3-connected planar graphs.
\newblock {\em International Journal of Computational Geometry and
  Applications}, 7(3):211--223, 1997.

\bibitem{cp-ltadp-95}
M.~Chrobak and T.~Payne.
\newblock A linear-time algorithm for drawing planar graphs.
\newblock {\em Inform. Process. Lett.}, 54:241--246, 1995.

\bibitem{fpp-sssfe-88}
H.~de~Fraysseix, J.~Pach, and R.~Pollack.
\newblock Small sets supporting {Fary} embeddings of planar graphs.
\newblock In {\em Proceedings of the 20th Annual ACM Symposium on Theory of
  Computing (STOC)}, pages 426--433, 1988.

\bibitem{harel98algorithm}
D.~Harel and M.~Sardas.
\newblock An algorithm for straight-line drawing of planar graphs.
\newblock {\em Algorithmica}, 20(2):119--135, 1998.

\bibitem{ht-ept-74}
J.~Hopcroft and R.~E. Tarjan.
\newblock Efficient planarity testing.
\newblock {\em Journal of the ACM}, 21(4):549--568, 1974.

\bibitem{k-dpguc-96}
G.~Kant.
\newblock Drawing planar graphs using the canonical ordering.
\newblock {\em Algorithmica}, 16:4--32, 1996.
\newblock (special issue on Graph Drawing, edited by G. {Di Battista} and R.
  Tamassia).

\bibitem{nakano4connected01}
K.~Miura, S.-I. Nakano, and T.~Nishizeki.
\newblock Grid drawings of 4-connected plane graphs.
\newblock {\em Discrete and Computational Geometry}, 26(1):73--87, 2001.

\bibitem{s-epgg-90}
W.~Schnyder.
\newblock Embedding planar graphs on the grid.
\newblock In {\em Proceedings of the 1st ACM-SIAM Symposium on Discrete
  Algorithms (SODA)}, pages 138--148, 1990.

\bibitem{t-hdg-63}
W.~T. Tutte.
\newblock How to draw a graph.
\newblock {\em Proceedings London Mathematical Society}, 13(52):743--768, 1963.

\end{thebibliography}
}

\newpage
\appendix
\noindent{\large \bf Appendix}
\thispagestyle{empty}

\setcounter{lemma}{2}
\begin{lemma} 
Let $C_{k}=(u=c_1, c_2, ..., c_r=v)$ be the exterior face of $G_k$
after the $k^{th}$ placement step. Let $\alpha(c_j, c_{j+1})$ denote
the angle of the vector $\vec{c_jc_{j+1}}$, for $1\leq j\leq r-1$. The
following holds for $2\leq k\leq m-1$:
\begin{enumerate}

\item 
$\alpha(c_j, c_{j+1})$ lies in $[-45^{\circ}, \arctan -1/2]
\cup \{0\} \cup [45^{\circ}, 90^{\circ}]$. It can not lie in $(-45^{\circ}, \arctan -1/2]$ if $c_j$ has a neighbor in $G-G_k$.

\item 
If $c_j\in C_{k}, c_j\notin \{c_1, c_r\}$ s.t. $c_j$ does not have a
neighbor in $G-G_k$, then:
\begin{enumerate}
\item If $Low(c_{j-1}, c_j)=j-1$ then $\alpha(c_j, c_{j+1})=90^{\circ}$ otherwise 
$\alpha(c_{j-1}, c_j)=-45^{\circ}$.
\item If $\alpha(c_j, c_{j+1})=90^{\circ}$ then $\alpha(c_{j-1}, c_j)\neq 90^{\circ}$.
\item If $\alpha(c_j, c_{j+1})=-45^{\circ}$ then $\alpha(c_{j-1}, c_j)\neq -45^{\circ}$. 
\end{enumerate} 
\end{enumerate}
\end{lemma}
\begin{Proof}

$Part-1.$\hspace{.2cm} We prove (1) by induction on $k$. For $k=2$, the lemma holds by the
placement of $u=c_1$, $w=c_2$ and $v=c_3$. Assume (1) holds for $k'=k-1$ where
$2\leq k'< m-1$.  Let $c_p, c_q \in C_{k-1}$, respectively be the
first and the last neighbor of the vertex(vertices) mapped to
$v_k$. If $|v_k|=1$ and $z_0$ is the vertex mapped to $v_k$, the
newly added edges on $C_k$, are $(c_p, z_0)$ and $(z_0, c_q)$. The
lemma holds for these edges by the placement algorithm. It holds for
the rest of the edges of $C_k$, except for $(c_q, c_{q+1})$, by
induction. For $(c_q, c_{q+1})$, if $z_0, c_q$ and $c_{q+1}$ are
aligned and $c_q$ does not have a neighbor in $G-G_k$, the placement
algorithm guarantees that $\alpha(c_q, c_{q+1})$ lies in
$(-45^{\circ}, \arctan -1/2]$, otherwise it holds by induction. For
$|v_k|=2$, (1) holds trivially by the placement algorithm, for the new
edges and by induction for the rest.

\vspace{.2cm}
$Part-2.$\hspace{.2cm} The proof of (2) is similarly by induction on $k$. For $k=2$, each of
$u=c_1$, $w=c_2$ and $v=c_3$ have neighbors in $G-G_k$ and the lemma holds
trivially. Assume (2) holds for $k'=k-1$ where $2\leq k'< m-1$ and
let $c_p, c_q \in C_{k-1}$ be defined as before.

We assume $|v_k|=1$. The case where $|v_k|=2$ is similar. Let $z_0$ be
the vertex mapped to $v_k$.  We need to prove that (2) holds for any
$c_j\in C_k=(c_1, ..., c_p, z_0, c_q, ..., c_r)$. Since by the
definition of canonical ordering, $z_0$ has an edge in $G-G_k$, (2)
holds for $c_j=z_0$. It holds for $c_j\in \{c_1, ..., c_{p-1}\}$ by
induction, since we do not make any changes in the locations of the
vertices in $\{c_1, ..., c_p\}$ after inserting $z_0$. It also holds
for $c_j \in \{c_{q+2}, ..., c_r\}$ by induction, since those vertices
are shifted by the same amount to the right after inserting
$z_0$. Then, we just need to prove that it holds for any $c_j\in
\{c_p, c_q, c_{q+1}\}$. We prove the cases where $c_j$ does not have
an edge in $G-G_k$, since otherwise the lemma holds trivially.
 
For $c_j=c_p$, we can safely assume that $c_{p-1}$ has at least one
neighbor in $G-G_k$.  This is true, since otherwise both $c_p$ and
$c_{p-1}$ have no neighbors in $G-G_k$. Because of the fact that $G$
is fully quadrilateralated this implies that $k=m$, which contradicts
the initial assumption about $k$. Now if $Low(c_{p-1}, c_p)=p-1$ then
(2a) holds trivially by the placement algorithm. If $Low(c_{p-1},
c_p)=p$, then by the first part of the lemma $\alpha(c_{p-1}, c_p)$
lies in $\{-45^{\circ}, 0\}\cup [45^{\circ}, 90^{\circ}]$.  The
placement algorithm guarantees that if $Low(c_{p-1}, c_p)=p$, then
$y(c_{p-1})>y(c_p)$. This implies $\alpha(c_{p-1}, c_p)=-45$.  The
proof of (2b) is by contradiction. Assume $\alpha(c_p,
z_0)=90^{\circ}$ and $\alpha(c_{p-1}, c_p)=90^{\circ}$ which, by the
placement algorithm, implies that $c_{p-1}$ doesn't have any neighbors
in $G-G_k$. Since both $c_p$ and $c_{p-1}$ don't have any edges in
$G-G_k$, $k=m$, which is a contradiction. (2c) holds by the placement
algorithm for $c_j=c_p$, since $\alpha(c_p, z_0)=90^{\circ}$.

For $c_j=c_q$, (2a), (2b) and (2c) hold by the placement 
algorithm. 

For $c_j=c_{q+1}$, if $c_q$ and $c_{q+1}$ are not aligned with $z_0$
during the initial placement, or $c_q$ has a neighbor in $G-G_k$, then
$c_q$ and $c_{q+1}$ are shifted the same amount to the right. Then in
this case, the lemma holds by induction. Note that we are assuming
that $c_{q+1}$ doesn't have a neighbor in $G-G_k$. Now assume that
after the initial placement all three are aligned and $c_q$ doesn't
have a neighbor in $G-G_k$ either. Since $G$ is $FQ$, this implies $k=m$, 
which contradicts the initial assumption about $k$.
\end{Proof}

\end{document}